\documentclass[journal,10pt,twoside]{IEEEtran}

\usepackage{amsmath, amssymb, amsfonts}
\usepackage{bm}
\usepackage{textcomp}
\usepackage{xcolor}
\usepackage{graphicx}
\usepackage{microtype}
\usepackage{booktabs}
\usepackage{cite}
\usepackage{algorithm}
\usepackage{algorithmic}
\usepackage{xcolor}
\usepackage{soul}
\usepackage{todonotes}
\usepackage{amsfonts}
\usepackage{multirow}
\usepackage{makecell}
\usepackage[T1]{fontenc}

\usepackage[
  colorlinks=true,
  linkcolor=blue!60!black,
  citecolor=blue!60!black,
  urlcolor=blue!60!black
]{hyperref}

\begin{document}

\title{Memory-- and Bandwidth--Efficient SPAD-LiDAR Ranging via Coarse-to-Fine Spline Sketching}

\author{Zhenya~Zang,~\IEEEmembership{}
        Istvan~Gyongy$^{\dagger}$,~\IEEEmembership{Senior Member,~IEEE,}
        ~and Mike~Davies$^{\dagger}$,~\IEEEmembership{Fellow,~IEEE}
\thanks{$^{\dagger}$These authors jointly supervised this work and 
        are co-senior authors.
        The authors are with the School of Engineering,
        University of Edinburgh, Edinburgh EH9 3FF, U.K.
        (e-mail: zzang@ed.ac.uk; igyongy2@exseed.ed.ac.uk; Mike.Davies@ed.ac.uk).} 
}

\maketitle

\begin{abstract}
This work presents an optimized compression framework for direct time-of-flight (dToF) light detection and ranging (LiDAR), using an accurate, compact timestamp-encoding strategy to address the high data-rate bottleneck from single-photon timing information in single-photon avalanche diode (SPAD) arrays. We propose a hardware-friendly timestamp-to-depth framework in which a sparse single-photon encoding strategy, namely coarse-to-fine spline sketches (CFSS), projects photon timestamps into fine-grained, low-dimensional representations, called sketch values. CFSS converts timestamps on-the-fly to sketch values and accumulates them with lower memory consumption, without constructing histograms, thereby saving on-chip/device memory. CFSS's closed-form solution for ToF retrieval ensures iteration-free reconstruction. Compared with the conventional sketched-LiDAR framework, the proposed method retains the same compression ratio in the single-peak case, ranging from hundreds$\times$ to thousands$\times$ depending on the accuracy–complexity trade-off, while improving depth-estimation accuracy. Both synthetic and real datasets are used to validate the performance. The CFSS framework is further extended to two-peak scenarios, enabling recovery of objects obscured by partially occluding camouflage and uniformly scattering semi-transparent media. Firmware implementation is also investigated by clearly separating online hardware processing from offline software workloads.
\end{abstract}

\begin{IEEEkeywords}
Light Detection and Ranging (LiDAR),
time-correlated single-photon counting (TCSPC),
Histogram-free compression,
single-photon avalanche diode (SPAD).
\end{IEEEkeywords}

\section{Introduction}
\label{sec:intro}
Single-photon detector-based direct time-of-flight (dToF) LiDAR is an emerging 3D sensing technology for robotic and autonomous systems that measures the ToF of photons from a light emitter to a receiver using time-correlated single-photon counting (TCSPC). Single-photon avalanche diode (SPAD) arrays \cite{istvan_4x4, hutchings2019, zhang201830, henderson2019192} are the primary candidate detectors due to their parallel sensing and single-photon sensitivity \cite{shin2016photon, tachella2019real}. Advances in semiconductor technology and sensor-architecture design enable higher lateral resolution and finer spatial information. A high-spatial-resolution, solid-state SPAD array avoids scanning, lowers cost, and mitigates mechanical vibration. 

However, the data throughput, typically the data rate at which ToF information from each photon is transferred, theoretically increases quadratically as the lateral resolution grows. At the system level, where multi-sensor and single-photon LiDAR fusion \cite{lindell2018single, sun2020spadnet} is widely used, LiDAR accounts for the largest share of data bandwidth because of its high data volume, including both spatial and temporal information. Multiple approaches address these high-data-rate challenges. The first group is on-chip partial histogramming, including binning \cite{binning}, sliding \cite{sliding}, and zooming \cite{zooming}, which has been reported to compress histograms across the full temporal dimension. A runtime peak-tracking method \cite{istvan_4x4} compresses per-pixel storage to only 8 histogram bins via a time gate that each pixel autonomously slides, locks onto the return peak, and identifies it. Equi-depth methods adaptively position histogram boundaries according to the photon-arrival distribution, yielding finer temporal resolution near the peak. Equi-depth histograms \cite{ingle2023_equi_depth} hierarchically estimate quantiles of the photon-arrival distribution on the fly, ensuring each bin contains approximately the same photon mass. Proportional equi-depth histograms \cite{sadekar2024eui_depth} instead employ independent proportional binners, each tracking a prescribed quantile concurrently over the exposure. Their reconstruction pipelines primarily focus on single-depth estimation. They report approximately \(10\text{--}100\times\) bandwidth reduction and validate their methods through hardware emulation. The second category is deep learning (DL), where an encoder compresses the histogram on-chip into latent space and an offline decoder reconstructs it \cite{Histless_lidar_auto}. Spiking deep learning for SPADs \cite{lin2024spiking, macleanTDCless} bypasses time-to-digital converters (TDCs) by processing SPAD events directly, thereby avoiding histogram construction. However, such deep-learning (DL) approaches are typically setup-dependent and exhibit limited mathematical interpretability. The third group is learning-free, statistics-driven, histogram-free compression. The representative study Sketched-LiDAR~\cite{sketch_lidar} projects per-pixel timestamps onto Fourier basis functions to compress them into a short vector (e.g., 4 to 32), and the hardware (HW)-friendly version of linear spline sketches \cite{spline_sketch} with fast closed-form (CF) ToF reconstruction, and the firmware implementation \cite{spline_fpga} on a field-programmable gate array (FPGA) to compute the sketch coefficients from timestamp streams for post-reconstruction. Sketched-LiDAR methods could be improved through adaptive peak identification, finer sketch allocation, and handling multi-peaks. Another histogram-free work \cite{Histless_lidar} emulates a deadtime-free SPAD response across multiple exposures and extracts the ToF from a single running mean of timestamps. 

As statistics-inspired Sketched-LiDAR methods offer mathematical interpretability and high reconstruction accuracy, we build on this line of work and demonstrate that sketch functions can be dynamically allocated at runtime to achieve finer temporal encoding resolution. We further extend the framework to two-peak conditions, which commonly arise in practical LiDAR applications. The contributions of this work are:
\begin{itemize}
    \item We optimize the hardware-efficient Sketched-LiDAR framework \cite{spline_sketch} by introducing a two-stage coarse-to-fine spline-sketch (CFSS) scheme. The proposed method first identifies the peak position using coarse sketches (CS), and then applies fine sketches (FS) with smaller intervals within the selected CS region. This enables custom CF solutions to extract finer relative-timing information without increasing per-photon arithmetic complexity, at the cost of modest additional LUT/BRAM storage on the firmware, compared to the original CS version \cite{spline_sketch}.

    \item We generalize the CFSS framework to two-peak scenarios by considering different peak occurrences. When two peaks fall within the same CS region, we use a shared-CS estimation strategy; when they fall in different CS regions, we apply two independently allocated FS sketches. The workload is clearly scheduled between the runtime firmware and post-processing software.

    \item We extensively evaluate the proposed CFSS framework on synthetic datasets under multiple evaluation conditions, including signal-to-background ratio (SBR) and instrument response function (IRF) full width at half maximum (FWHM). Real datasets with one- and two-peak signals from partially occluding and semi-transparent media, acquired at different photon-counting levels, are used for evaluation with varying numbers of sketch coefficients.
\end{itemize}

\section{Problem Definition}
\label{sec:problem}
\begin{table}[!t]
\centering
\scriptsize
\caption{Coarse-sketch (CS) and fine-sketch (FS) variables.}
\label{tab:coarse_vs_fine}
\renewcommand{\arraystretch}{1.2}
\begin{tabular}{lll}
\toprule
& \multicolumn{1}{c}{\textbf{CS}}
& \multicolumn{1}{c}{\textbf{FS}} \\
\midrule
Photons
& $\{x_j\}_{j=1}^{n_{\mathrm C}}$ on $[0,T)$
& $\{x_j\in\mathcal W_\ell\}_{j=1}^{n_{\mathrm{F},W}}$ \\
Sketch
& $\hat{\bm z}$
& $\hat{\tilde{\bm z}}$ \\
Spacing
& $\Delta=T/M$
& $\tilde\Delta=W/M$ \\
Resolution gain
& ---
& $\Delta/\tilde\Delta=M/w_h$ \\
SBR
& $\mathrm{SBR}$
& $(T/W)$ \\
Count
& $n_{\mathrm C}$
& $n_{\mathrm{F},W}$ \\
Peak bin
& $\ell=\arg\max_j \hat z_j$
& $\tilde\ell=\arg\max_j \hat{\tilde z}_j$ \\
Depth
& $\hat t_{\rm coarse}$
& $\hat t_{\rm fine}$ \\
\bottomrule
\end{tabular}
\end{table}

\begin{table}[!t]
\centering
\caption{Three-scenario CF depth estimator used in both CS and FS \cite{spline_sketch}.}
\label{tab:cf_scenarios}
\renewcommand{\arraystretch}{1}
\begin{tabular}{ll}
\toprule
\multicolumn{1}{c}{\textbf{Scenario}}
& \multicolumn{1}{c}{\textbf{Estimate}} \\
\midrule
1 &
$\hat t^{(1)}
=\xi_\ell+\frac{\Delta}{2}
+\frac{\Delta(z_\ell-z_{\ell-1})}{2\hat\alpha_1}$ \\
2 &
$\hat t^{(2)}
=\xi_{\ell+1}+\frac{\Delta}{2}
+\frac{\Delta(z_{\ell+1}-z_\ell)}{2\hat\alpha_1}$ \\
3 &
$\hat t^{(3)}
=\xi_{\ell+1}
+\frac{\Delta(z_{\ell+1}-z_{\ell-1})}{\hat\alpha_1}$ \\
Selection &
$s=\arg\min_k\|\bm z-\mathbb E_\pi\bm\Phi_1\|^2$ \\
Output &
$\hat t=\hat t^{(s)}$ \\
\bottomrule
\end{tabular}

\vspace{0.3em}
{\footnotesize
$\bm z$ denotes the corresponding sketch values of CS or FS shown in (\ref{eq:cs-sketch}) and (\ref{eq:zs-sketch}).}
\end{table}

\subsection{Observation Model}
\label{sec:obs_model}
A SPAD records $n$ photon timestamps $\{x_j\}_{j=1}^{n} \subset [0,T)$ over
$N$ laser pulses, where $T$ is the number of time bins. For a scene containing
a single surface at depth $d^{\star}$ with ToF $t^{\star}=2d^{\star}/c$, each $x_j$ is independently sampled
\begin{equation}
    \pi(x_j \mid t^{\star},\alpha_1)
    =
    \alpha_1 \pi_s(x_j \mid t^{\star})
    +
    \alpha_0 \pi_b(x_j),
    \label{eq:fm-mixture}
\end{equation}
where $\pi_b(x)=1/T$ denotes uniform background arrivals and
$\alpha_0=1-\alpha_1=1/(1+\mathrm{SBR})$. The echo signal is modeled as the convolution of a Dirac impulse at the target delay \(t^*\) with the system's IRF, under an idealized detector model that neglects imperfection effects such as dead time and pile-up
\begin{equation}
    \pi_s(x\mid t^{\star})
    =(\delta(x-t^{\star}) * h(x))/{H},
\end{equation}
where $h$ is the IRF, modeled as a Gaussian with FWHM, and $H=\sum_{t=0}^{T-1} h(t)$ is the normalization factor. The objective is to estimate $t^{\star}$ from timestamps $\{x_j\}$.

\subsection{Coarse Sketch (CS)}
\label{sec:coarse_sketch}
CFSS follows the general principle of adaptive temporal setting: a low-resolution representation first identifies the region containing a likely peak, after which finer representation capacity is concentrated within that region. This is similar to sliding, zooming, and peak-tracking histograms, where temporal bins are dynamically reallocated around an estimated return. We apply this principle in the sketch domain. The CS covers the full measurement range and primarily localizes returns, while the FS reallocates the same number of sketch channels over a finer temporal support for finer resolution. Spline-sketched LiDAR \cite{spline_sketch} avoids constructing histograms by projecting timestamps using a linear spline sketch with $ M$ coefficients, $M \ll T$, where $M$ is the number of sketches. Let $\xi_0, \dots, \xi_{M-1}$ be the equispaced knots on $[0,T]$ with knot spacing $\Delta = T/M$. The $i$-th basis function and the empirical sketch are
\begin{equation}
    \phi_{i,1}(x) = \phi_1\!\bigl( (x \bmod T)/\Delta - i\bigr),
    \qquad i = 0,\dots,M-1,
    \label{eq:cs-basis}
\end{equation}
\begin{equation}
    \hat{\mathbf z}
    =
    \frac{1}{n_{\mathrm{C}}}
    \sum_{j=1}^{n_{\mathrm{C}}}
    \mathbf{\Phi}_1(x_j^{\mathrm C})
    \quad
    \bm{\Phi}_1(x) = [\phi_{i,1}(x)]_{i=0}^{M-1},
    \label{eq:cs-sketch}
\end{equation}
with $\phi_1$ the cardinal linear B-spline \cite{spline_sketch}; $\phi_{i,1}$ peaks at $\xi_{i+1}$ and is supported on $[\xi_i,\xi_{i+2}) \bmod T$. We develop the proposed method for $p=1$; a $p=0$ zero-order variant, equivalent to coarse binning \cite{spline_sketch}, serves as a baseline. Given $\hat{\bm{z}}$, the CF solution of \cite{spline_sketch} first identifies the winning bin $\ell = \arg\max_j \hat z_j$ and estimates the signal fraction $\hat\alpha_1$ from the background coefficients. It then computes three ToF values of the CF candidate (Eqs.~(11)--(13) of
\cite{spline_sketch}) and returns $\hat t$ to minimize
$\|\hat{\bm{z}} - \mathbb{E}_\pi \bm{\Phi}_1\|_2^2$, from which the depth is calculated $\hat d = c\,\hat t / 2$.

Notably, CFSS is a two-stage process rather than two full acquisitions. For each exposure, a short CS acquisition first identifies the approximate peak's location, after which the system switches to FS for the remaining acquisition to refine the depth estimate. The system refreshes the CS and FS for every new exposure, without relying on the previous frame. Thus, CFSS uses part of each exposure for CS and the remaining part for FS, rather than requiring two full exposures. The CS stage could potentially be replaced by in-pixel zooming histogramming \cite{istvan_4x4}, with the identified zoomed region then used to define the FS; photons accumulated during the zooming stage could be reused for the FS. In our experiments, however, we consider the two-stage CFSS scheme without assuming histogram zooming, and use a fixed total acquisition budget with a configurable ratio controlling the fraction of photons assigned to the CS and FS stages, corresponding in practice to the fraction of acquired frames used by each stage. Let $\rho$ denote the fraction of the total photons allocated to the CS, such that $n_{\mathrm{C}}=\rho n$ and
$n_{\mathrm{F}}=(1-\rho)n$ photons are assigned to the CS and FS stages. We denote the CS and FS photon sets by $\{x_j^{\mathrm C}\}_{j=1}^{n_{\mathrm C}}$ and
$\{x_j^{\mathrm F}\}_{j=1}^{n_{\mathrm F}}$. We use $\rho=0.1$ in all experiments, allocating $10\%$ of the total photons to CS and $90\%$ to FS.

\subsection{Fine Sketch (FS)}
\label{sec:fine_sketch}
\begin{figure}[!t]
    \centering
    \includegraphics[width=\linewidth]{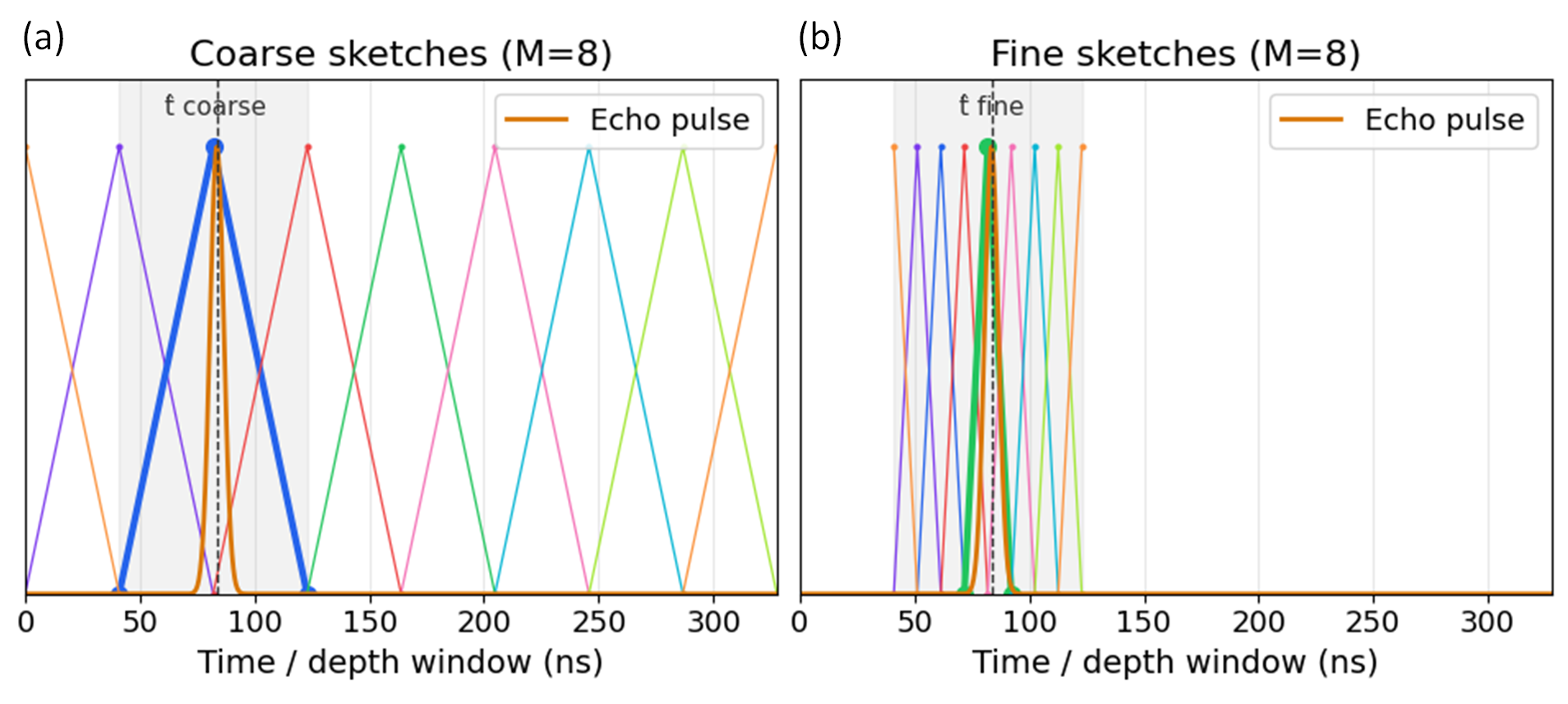}
    \caption{Coarse and fine sketches for $M=8$, $n=500$,
    IRF FWHM = 1 ns, $T = 4096$, histogram bin width is 80 ps, SBR $=5$.
    (a)~Coarse sketches (CS): $M$ bases $\{\phi_{i,1}\}$ over $[0,T)$; winning basis $\phi_{\ell,1}$ in bold, estimate $\hat t_{\mathrm{coarse}}$ dashed. (b)~Fine sketches (FS): $M$ bases of local spacing $\tilde\Delta = W/M$ are allocated to $\mathcal{W}_\ell$ (grey) of width $W = 2\Delta$. The echo pulses are normalized to align with the sketch's amplitude for better visualization.}
    \label{fig:basis_zoom}
\end{figure}
Fig. \ref{fig:basis_zoom} illustrates an example of the CS-to-FS principle, with several representative parameters. The FS is a second refinement stage applied to the CS. With $\ell = \arg\max_j \hat z_j$ obtained from the CS over the entire measurement range, we define a fine window of width $W = w_h \Delta$ ($w_h = 2$) centered on the basis peak $\xi_{\ell+1}$
\begin{equation}
    \mathcal{W}_\ell
    \;=\; \bigl[\, w_{\mathrm{lo}},\; w_{\mathrm{lo}} + W \,\bigr] \bmod T,
    \quad
    w_{\mathrm{lo}} = \xi_{\ell+1} - W/2.
    \label{eq:zs-window}
\end{equation}
$w_h = 2$ recovers the full support of $\phi_{\ell,1}$. If the window crosses the boundary of the measurement range, it wraps between $T$ and $0$ rather than being truncated, following the periodic convention commonly used for TCSPC histograms over one laser repetition period. The FSs use $M$ bases of new spacing $\tilde\Delta = W/M$ located within local knots $\tilde\xi_i = w_{\mathrm{lo}} + i\tilde\Delta$ ($i = 0,\dots,M$), with period $\mathcal{W}_\ell$:
\begin{equation}
    \tilde\phi_{i,1}(x)
    \;=\; \phi_1\!\Bigl(\tfrac{x - \tilde\xi_i}{\tilde\Delta}\Bigr),
    \qquad x \in \mathcal{W}_\ell.
    \label{eq:zs-basis}
\end{equation}
The coefficient of the FS is computed directly on $x$ within $W$
\begin{equation}
    \hat{\tilde{\mathbf z}}
    =
    \frac{1}{n_{\mathrm{F},W}}
    \sum_{j:\,x_j^{\mathrm{F}}\in\mathcal{W}_{\ell}}
    \tilde{\mathbf{\Phi}}_1(x_j^{\mathrm{F}}),
    \qquad
    n_{\mathrm{F},W}
    =
    \left|
    \{j:x_j^{\mathrm{F}}\in\mathcal{W}_{\ell}\}
    \right|.
    \label{eq:zs-sketch}
\end{equation}
which runs on firmware (FPGA). The conditional distribution of $x \in \mathcal{W}_\ell$ retains
the form of \eqref{eq:fm-mixture}
\begin{equation}
    \tilde\pi(x \mid t^\star, \tilde\alpha_1;\,\mathcal{W}_\ell)
    \;=\; \tilde\alpha_1\, \pi_s(x \mid t^\star)
        \;+\; (1 - \tilde\alpha_1)\,/\,W,
    \label{eq:zs-mixture}
\end{equation}
with the same Gaussian signal density $\pi_s$ and the same IRF FWHM. The FS scheme differs from the CS scheme \eqref{eq:fm-mixture} in two ways: the density is $1/W$ (uniform on $\mathcal{W}_\ell$) rather than $1/T$, and the in-window signal fraction $\tilde\alpha_1$ is estimated by the same background-bin construction as the CS \cite{spline_sketch} applied to $\hat{\tilde{\bm z}}$. Unlike the CS, which considers background from all bins, the FSs process only the $W/T$ fraction of the background, enhancing in-window SBR and sketch-domain resolution.

\begin{figure*}[!t]
    \centering
    \includegraphics[width=\textwidth]{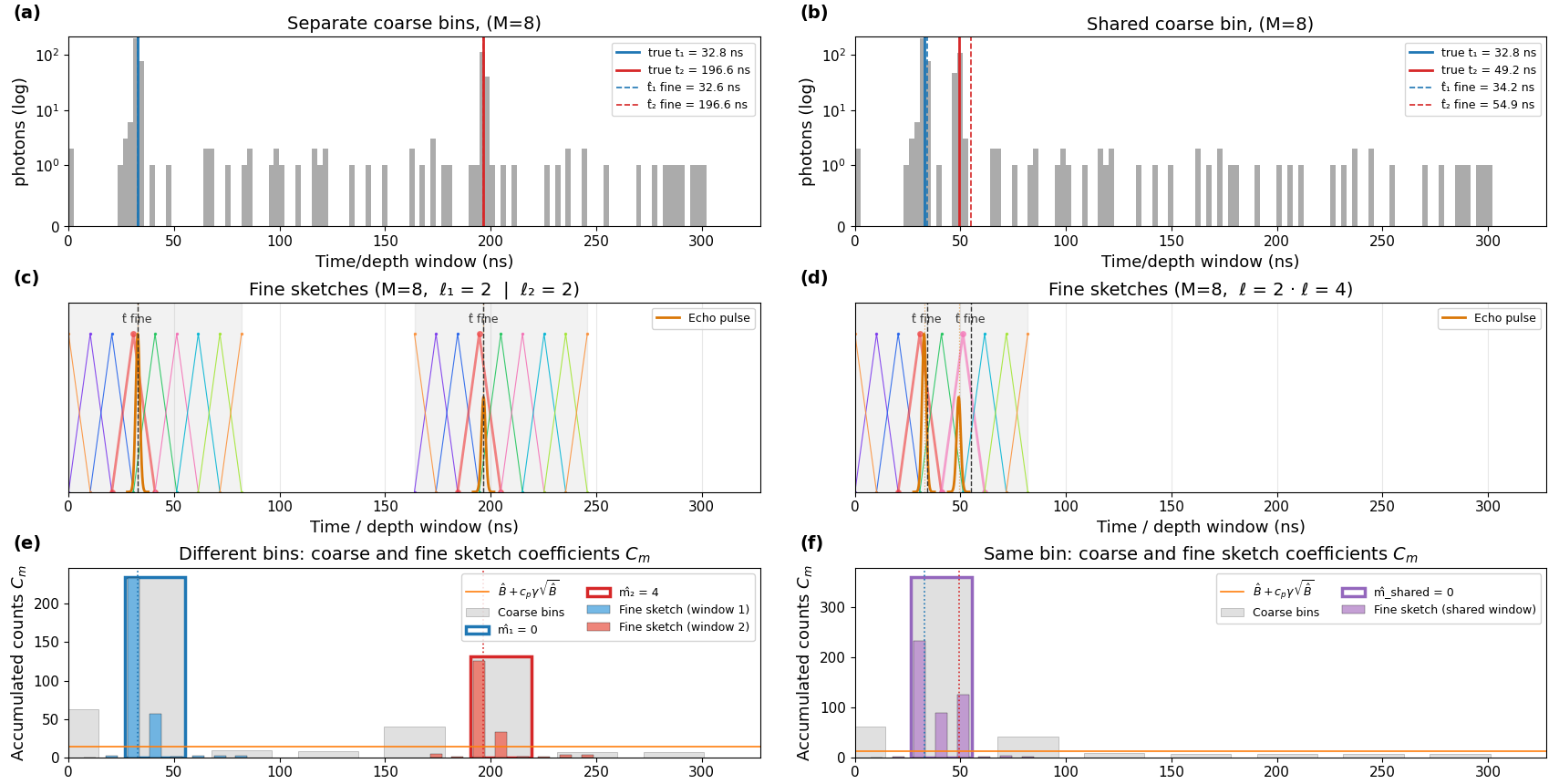}
    \caption{Two-peak detection using CFSS. The left column shows two returns
    detected in separate CS regions; the right column shows the case where they
    fall within a shared coarse region. (a),(b)~Photon histograms for the
    separate-bin and shared-bin cases; solid verticals mark the true
    $t_1, t_2$ and dashed verticals the fine estimates $\hat t_1, \hat t_2$.
    (c),(d)~FS basis functions after local window reallocation; orange curves
    show the normalized peak-1 and peak-2, scaled by the relative amplitude to
    peak-1. (e),(f)~Accumulated coarse and fine coefficients $C_m$ with the
    detection threshold $\hat B + \gamma\sqrt{c_p \hat B}$; dotted verticals mark
    the true $t_1, t_2$. In~(e), two independent fine windows are allocated;
    in~(f), a single shared window covers both peaks.}
    \label{fig:two_peak}
    \label{fig:two_peak_scenario}
\end{figure*}

As \eqref{eq:zs-mixture} is structurally identical to
\eqref{eq:fm-mixture}, the CF solution of~\cite{spline_sketch} applies to $\hat{\tilde{\bm z}}$ under the substitutions $\Delta \to \tilde\Delta$ and $\xi_i \to \tilde\xi_i$; the three-scenario CF arithmetic is unchanged. The only difference from CS is that the FPGA accumulates $\phi_{i,1}$ over the in-window timestamps $\{x_j: x_j \in \mathcal{W}_\ell\}$ rather than over the full measurement range. Like the CS decoder, the FS decoder operates in absolute coordinates. It adds the offset of the winning CS index to estimate the depth $\hat t_{\mathrm{fine}}$ directly, without rescaling. Table~\ref{tab:coarse_vs_fine} summarizes the two passes, where only the input sketch and the window-local parameters $(\tilde\Delta,\,\tilde\xi_i,\,n_{\mathrm{F},W})$ differ. Table \ref{tab:cf_scenarios} presents the CF solutions of CS and FS.

In particular, the CF solution for CS ToF reconstruction assumes $FWHM\le\Delta$, allowing large $M$ (with finer FS) when the IRF is narrow. The FS also requires the FS placement condition, meaning that $FWHM \le \tilde\Delta = w_h T / M^2$ becomes restrictive as $M$ grows. The criterion on $M$ for FS, derived from $FWHM \times M^2/(w_h T) \lesssim 1$, is
\begin{equation}
    M \;\lesssim\; \sqrt{\frac{w_h\,T}{FWHM}}.
    \label{eq:zs-regime}
\end{equation}

\subsection{Hardware Consideration}
\label{sec:HW}
The sketch function $\mathbf{\Phi}$ encoding timestamps to the sketch coefficient $\hat{\tilde{\bm z}}$ for the FS in \eqref{eq:zs-sketch} can be implemented as look-up tables (LUTs), precomputed offline using \eqref{eq:zs-basis} and loaded on-chip. The CS LUTs are precomputed in the same way using \eqref{eq:cs-basis}, with a different knot spacing and a separately instantiated set of on-chip LUTs. The LUTs eliminate the explicit on-chip computing. Because only $(\tilde\Delta, \tilde\xi_i, n_{\mathrm{F},W})$ differ from the coarse case, and $\tilde\Delta, \tilde\xi_i$ can also be precomputed offline, the only per-pixel runtime quantity is $n_{\mathrm{F},W}$, obtained by discarding photons that fall outside $\mathcal{W}_\ell$. Within each exposure, $\mathcal{W}_\ell$ can be identified by running CS mode on the initial frames to collect enough photons for a coarse estimate, retrieve the CS index that accommodates FS afterward, and then switch to FS mode for the remaining frames of the exposure. Taking the QuantiCAM \cite{henderson2019192} SPAD array as an example, which generates per-pixel timestamps for firmware processing, the sketch coefficient processing element processes each timestamp as $(x-\tilde\xi_i)/\tilde\Delta$, where the subtraction completes in one clock cycle. The division can be implemented as a shift operation since the number of time bins is usually an integer power of two. The computed sketch coefficient addresses $M$ LUTs of $\Phi$ in parallel to accumulate the $M$-component sketch coefficients \cite{spline_fpga} simultaneously. Therefore, the FPGA instantiates $M$ sets of LUTs. For example, with CS at $M = 4$, 16-bit entries, and 32-deep LUTs, the storage requirement is $4 \times 32 \times 16 / (8 \times 1024) = 0.25$~KB; FS adds another equivalent LUT set, giving $0.5$~KB in total, which is an acceptable LUT/BRAM consumption for FPGAs. Memory scales proportionally with $M$, and prior work \cite{spline_sketch} showed that $M = 10$ achieves high accuracy with only marginal improvement from larger $M$.

\subsection{Two Peak Scenario}
\label{sec:two_peak}
\begin{table}[!t]
\centering
\caption{HW--SW workload partition for FS processing.
CF denotes the three-scenario closed-form decoder.}
\label{tab:hw_sw_partition}
\scriptsize
\renewcommand{\arraystretch}{1.05}
\setlength{\tabcolsep}{3pt}
\begin{tabular}{@{}ccc@{}}
\toprule
\textbf{Case} & \textbf{Firmware / FPGA} & \textbf{Software} \\
\midrule
\makecell{\textbf{One peak} \\\textbf{Compression ratio: $T/M$}}
& \makecell{$\{x_j\}\!\to\!C_m$~\eqref{eq:two-peak-Cm}\\
$\{x_j\!\in\!\mathcal W_\ell\}\!\to\!\hat z$~\eqref{eq:zs-sketch}}
& \makecell{$\ell=\arg\max_m C_m$\\
$\hat B =\min_{m\in\mathcal{U}} C_m$ \\
$C_{\hat m_2}\leq\hat B+\gamma\sqrt{c_p\hat B}$\\
$\hat t_{\rm fine}={\rm CF}(\hat z)$} \\
\midrule
\makecell{\textbf{Two peaks,}\\\textbf{disjoint CS bins}\\\textbf{Compression ratio: $T/2M$}} 
& \makecell{$\{x_j\}\!\to\!C_m$~\eqref{eq:two-peak-Cm}\\
$\{x_j\!\in\!\mathcal W_{\hat m_q}\}\!\to\!\hat z_q$~\eqref{eq:zs-sketch}}
& \makecell{$\hat m_1=\arg\max_m C_m$\\
$\hat B =\min_{m\in\mathcal{U}} C_m$\\
$C_{\hat m_2}>\hat B+\gamma\sqrt{c_p\hat B}$\\
$\hat t_1={\rm CF}(\hat z_1)$\\
$\hat t_2={\rm CF}(\hat z_2)$} \\
\midrule
\makecell{\textbf{Two peaks,}\\\textbf{shared CS bin}\\\textbf{Compression ratio: $T/M$}}
& \makecell{$\{x_j\}\!\to\!C_m$~\eqref{eq:two-peak-Cm}\\
$\{x_j\!\in\!\mathcal W_{\hat m_1}\}\!\to\!\hat z$~\eqref{eq:zs-sketch}}
& \makecell{$\hat t_1^{(0)}={\rm CF}(\hat z)$\\
$r_z=\hat z-\hat\alpha_1\,g(\hat t_1^{(0)})$\\
$r_C = n_{\mathrm{F},W}r_z$ \\
$r_{C,\max}>\hat B'+\gamma\sqrt{c_p\hat B'}$\\
$\hat t_2={\rm CF}(r_z)$\\
$\hat t_1={\rm CF}\!\bigl(\hat z-\hat\alpha_2\,g(\hat t_2)\bigr)$} \\
\bottomrule
\end{tabular}
\end{table}
In practice, a SPAD pixel may receive photons from more
than one surface due to partial occlusion or multi-surface returns. For example, when a semi-transparent or partially occluding object or a scattering medium is located in front of a second surface, the illumination region may be partially blocked by the front object and partially transmitted to the background surface. Multiple returns can also sometimes arise at object boundaries, where a pixel observes surfaces at different depths, or from reflective multipath that generates ghost returns. We therefore extend the proposed FS framework to the two-peak conditions. The observation model becomes
\begin{equation}
    \pi(x_j \mid t_1^\star,t_2^\star,\alpha_1,\alpha_2)
    =
    \alpha_1 \pi_s(x_j \mid t_1^\star)
    +
    \alpha_2 \pi_s(x_j \mid t_2^\star)
    +
    \alpha_0 \pi_b(x_j),
    \label{eq:two-peak-mixture}
\end{equation}
where $t_1^\star$ and $t_2^\star$ denote the ToFs of the two returns, $\alpha_1$ and $\alpha_2$ denote their effective received return strengths, and $\alpha_0$ is the background fraction. In synthetic two-peak experiments, we assume the two returns differ in amplitude, as is often the case in practice. CS first computes the unnormalized sketch coefficients
\begin{equation}
    C_m=\sum_{j=1}^{n_{\mathrm C}}\phi_{m,1}(x_j^{\mathrm C}),
    \qquad m=0,\ldots,M-1.
    \label{eq:two-peak-Cm}
\end{equation}
rather than the normalized sketch in \eqref{eq:cs-sketch}. $C_m$ is an
accumulated sketch coefficient on a photon-count scale. The dominant return is detected as
\begin{equation}
    \hat m_1 = \arg\max_m C_m .
    \label{eq:two-peak-first}
\end{equation}
To detect peak-2, we first remove the neighborhood of the dominant CS peak. Since a return peak contributes not only to its own coefficient $C_{\hat m_1}$ but also to neighboring ones, the large $C_m$ around $\hat m_1$ may be leakage from peak-1. Let $\mathcal{U}$ denote the unmasked sketch bins after suppressing the neighborhood of the dominant peak. The masking does not impose a minimum separation on the overall two-peak framework: sufficiently separated returns are detected as distinct CS regions, whereas closer returns within the same CS region are handled by the shared-window FS procedure. We estimate the background level as $\hat B=\min_{m\in\mathcal U}C_m$ and calculate the maximum unmasked sketch coefficient,
\begin{equation}
    \hat m_2
    =
    \arg\max_{m\in\mathcal{U}} C_m .
    \label{eq:two-peak-second}
\end{equation}
Following the background thresholding rule used in the in-pixel peak
detector \cite{gnecchi20171}, the peak-2 is accepted only if
\begin{equation}
    C_{\hat m_2}
    >
    h_{\mathrm{thresh}}
    =
    \hat B + \gamma\sqrt{c_p\hat B}.
    \label{eq:two-peak-threshold}
\end{equation}
Notably, $\gamma=1.75$ was justified and used in \cite{istvan_4x4}; however, it applies to the histogram or coarse-binning (equivalently, zero-order spline sketches \cite{spline_sketch}), whereas we operate in the triangle-shaped (first-order) sketched domain. Because the first-order linear basis spreads each photon across two adjacent coefficients, background fluctuations between bins are smaller than in the histogram case, making the peak stand out. The Poisson variance per coefficient is therefore $(2/3)\hat B$ rather than $\hat B$\footnote{The factor follows from Poisson statistics. For a uniform photon flux, the variance of each sketch coefficient scales with the squared $L^2$ norm of the basis function, according to Campbell's theorem \cite{kingman1992poisson}. For the triangular $\phi_1(x) = x\,\mathbb{1}_{[0,1)}(x) + (2-x)\,\mathbb{1}_{[1,2)}(x)$\cite{spline_sketch}, $\int_0^2 \phi_1^2(u)\,\mathrm{d}u = \int_0^1 x^2\,\mathrm{d}x + \int_1^2 (2-x)^2\,\mathrm{d}x = \left[\tfrac{x^3}{3}\right]_0^1 + \left[-\tfrac{(2-x)^3}{3}\right]_1^2 = \tfrac{1}{3} + \tfrac{1}{3} = \tfrac{2}{3}$.}, and \eqref{eq:two-peak-threshold} is rescaled with $c_p=2/3$ accordingly. If \eqref{eq:two-peak-threshold} is satisfied, a second return is detected in a different CS region; otherwise, no separate coarse-region return is detected and the measurement is passed to the shared-window FS test.

For the different CS-bin case, two fine windows are allocated
independently, one around each detected peak. Specifically,
\begin{equation}
    \mathcal{W}_{\hat m_q}
    =
    \left[
    \xi_{\hat m_q+1}-W/2,\;
    \xi_{\hat m_q+1}+W/2
    \right]\bmod T,
    \qquad q\in\{1,2\},
    \label{eq:two-peak-two-windows}
\end{equation}
with $W=w_h\Delta$. Each window is processed using the same FS
construction \eqref{eq:zs-sketch}, and the CF estimator is applied
independently to obtain $\hat t_1$ and $\hat t_2$. This preserves the
same $M$ per peak while applying the second-window offset.

For the shared coarse-bin case, place one fine window at $\hat m_1$. The FS contains both returns. We first apply the CF estimator to obtain the dominant return. We then subtract the signal sketch with the largest sketch coefficient from the measured FS and apply the CF estimator again to the residual. This procedure recovers two close peaks that the CS stage cannot separate. If the estimated residual amplitude is below the threshold, retain the measurement as a single-peak estimate. 

\begin{figure}[t]
    \centering
    \includegraphics[width=\columnwidth]{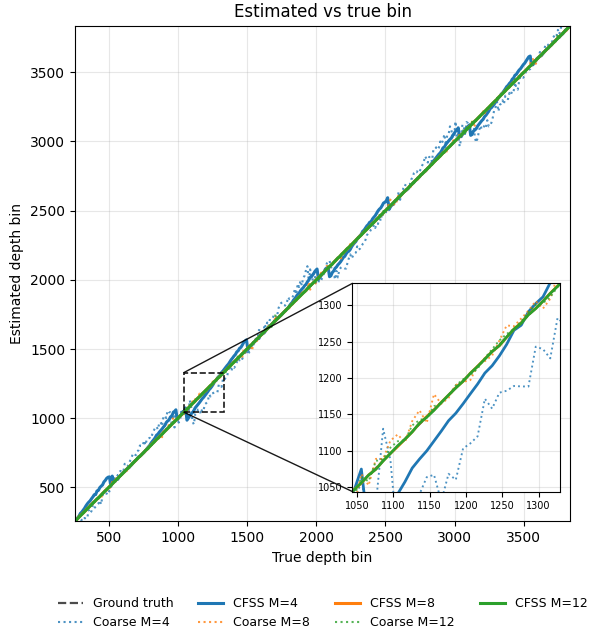}
    \caption{Estimated vs.\ true bin on synthetic data
        ($n{=}500$ photons, FWHM${=}2$\,ns, SBR${=}8$, $T{=}4096$ bins,
        bin width $80$\,ps). CS (dotted) and FS (solid) estimates are shown for
        $M \in \{4, 8, 12\}$. Each point is averaged over 50 trials at one true
        depth, and the zoomed insets show that the estimates approach the ground
        truth as $M$ increases.}
    \label{fig:depth_sweep}
\end{figure}

\begin{figure}[t]
    \centering
    \includegraphics[width=\linewidth]{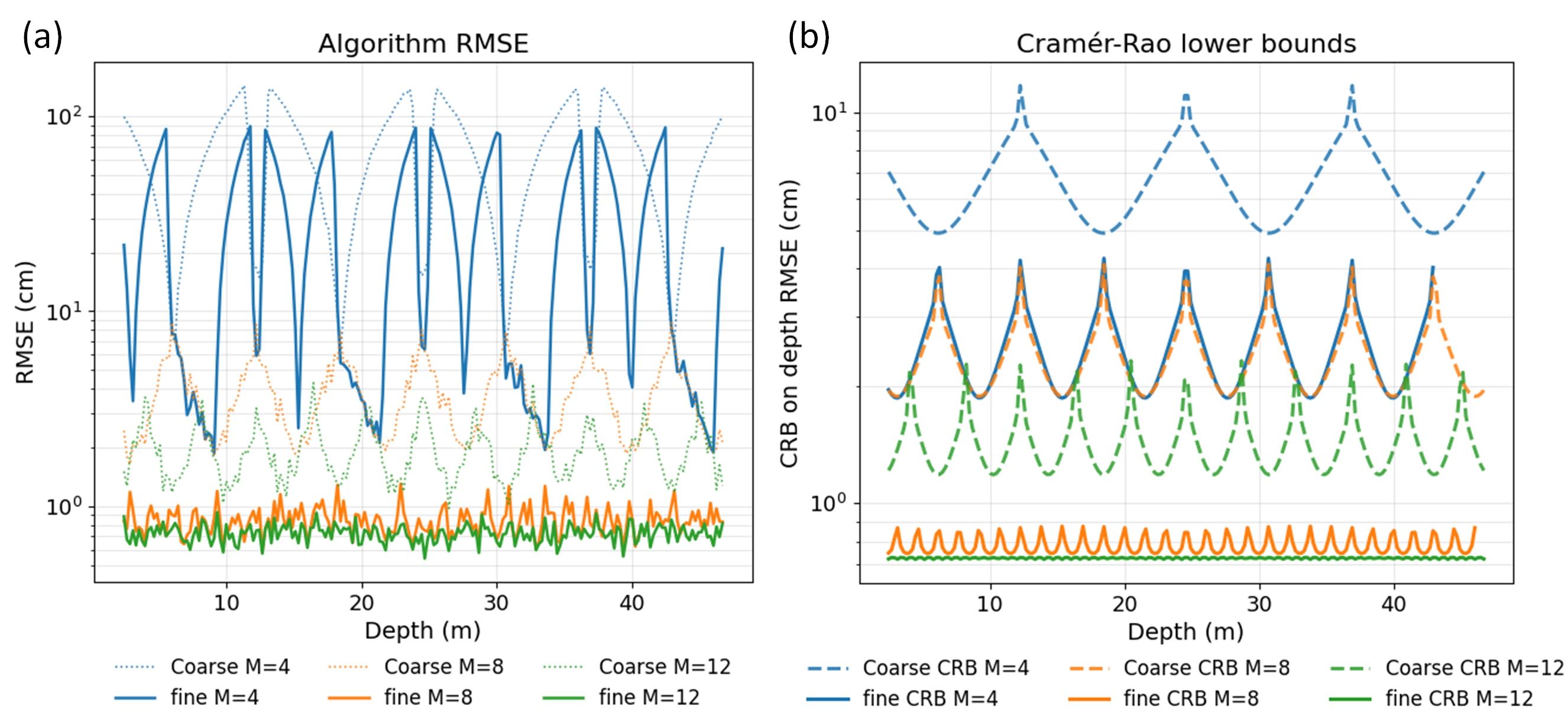}
    \caption{(a)~Per-depth RMSE over $50$ trials and
            (b)~Cram\'er--Rao lower bound (CRB) on the depth estimate, for CS (dotted) and FS (solid) sketches with $M \in \{4, 8, 12\}$. Configuration as in Fig.~\ref{fig:depth_sweep}; period $T = 4096$ bins corresponds to ${\approx}\,49$\,m range.}
    \label{fig:rmse_crb}
\end{figure}

\begin{figure}[t]
\centering
\includegraphics[width=\linewidth]{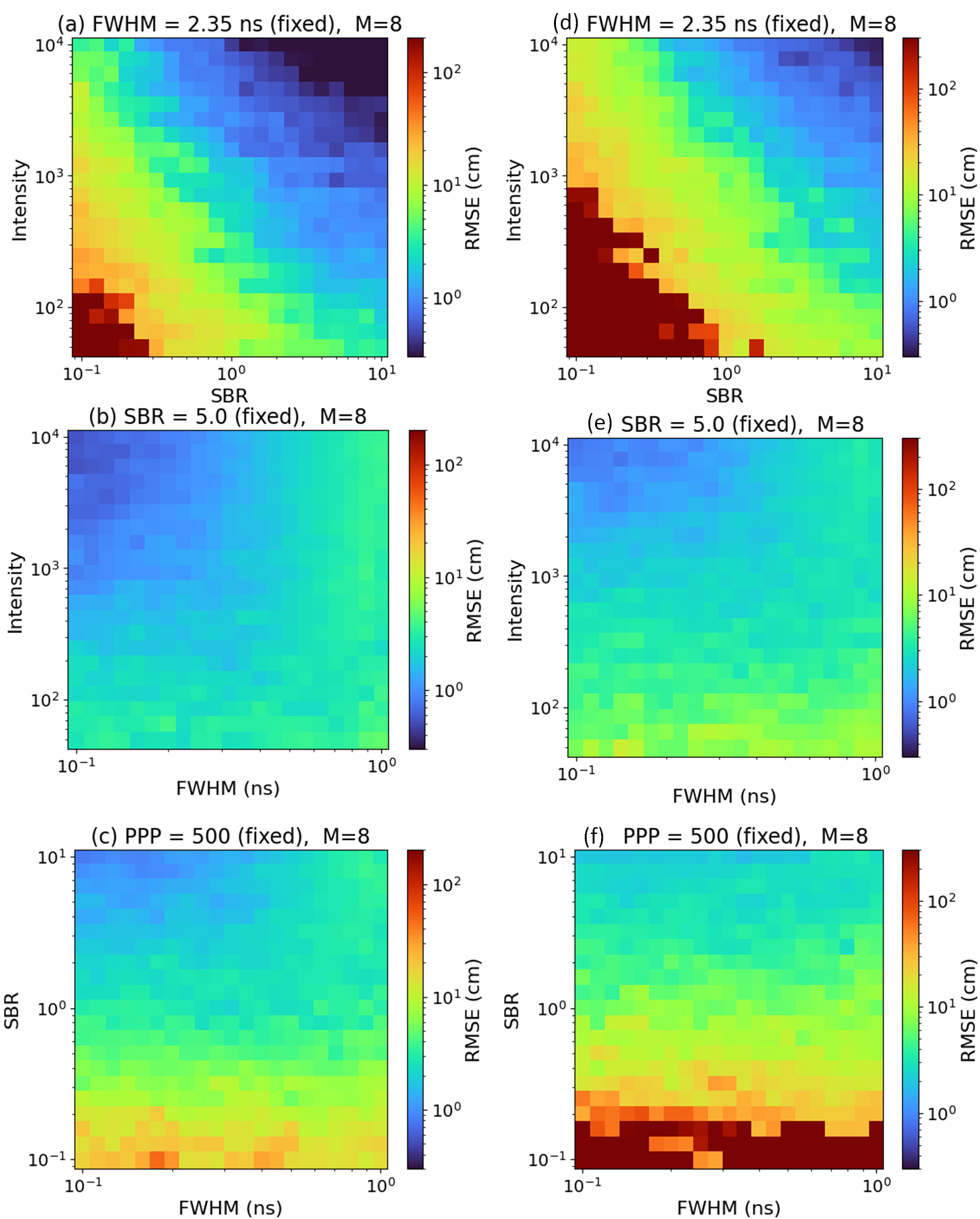}
\caption{Depth RMSE (cm) across FWHM, intensity, and SBR
at $M=8$, for the one-peak pipeline (a--c) and the two-peak
pipeline's second return (d--f).
(a),(d) Intensity vs.\ SBR at fixed FWHM $=2.35$~ns;
(b),(e) intensity vs.\ FWHM at fixed SBR $=5$;
(c),(f) SBR vs.\ FWHM at fixed intensity $=500$~photons/pixel (PPP).
Each cell averages 30 trials. The color scales $[0.3, 200]$~cm and $[0.3, 300]$~cm are for the one-peak and two-peak panels.}
\label{fig:heatmaps_onepeak}
\end{figure}

\begin{figure}[!t]
\centering
\includegraphics[width=\linewidth]{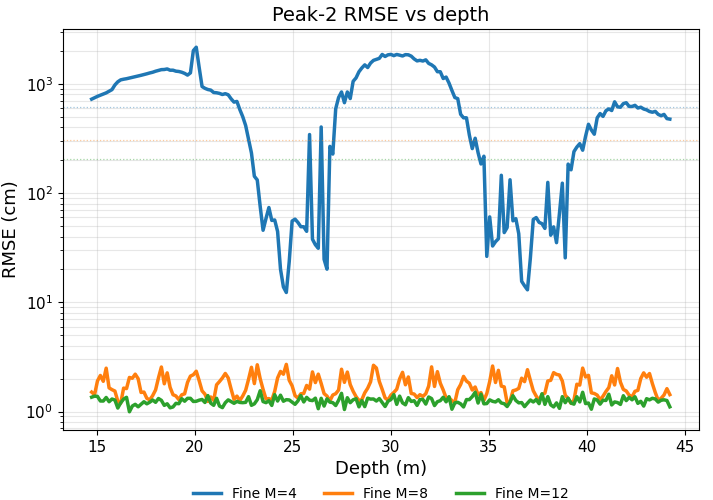}
\caption{Peak-2 depth RMSE versus depth. Peak-1 fixed at
$t_1{=}$4.9~m; peak-2 swept over different distances.
Parameters: $\alpha_1{=}0.6$, $\alpha_2{=}0.4$, FWHM = 2 ns,
SBR${=}8$, $n{=}500$ photons, 100 trials per depth.}
\label{fig:peak2_rmse_vs_depth}
\end{figure}

\begin{figure}[!t]
\centering
\includegraphics[width=\linewidth]{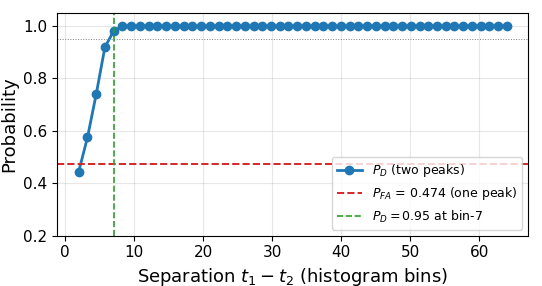}
\caption{Same-CS two-peak separability at $M=8$ as the peak distance
$|t_1 - t_2|$ varies, with peak-1 centered in its coarse basis.
$P_D$ (blue) is the fraction of 200 two-peak trials in which the residual
exceeds the threshold of~\eqref{eq:two-peak-threshold}, i.e. both returns are
detected. $P_{FA}$ (red dashed) is the false-alarm rate, i.e., the fraction
of one-true-peak trials where peak-2 is spuriously detected.
The green dashed line marks the point at which $P_D$ first reaches 95\%.}
\label{fig:two_peak_stress_test_M8}
\end{figure}

\begin{figure*}[t]
    \centering
    \includegraphics[width=\textwidth]{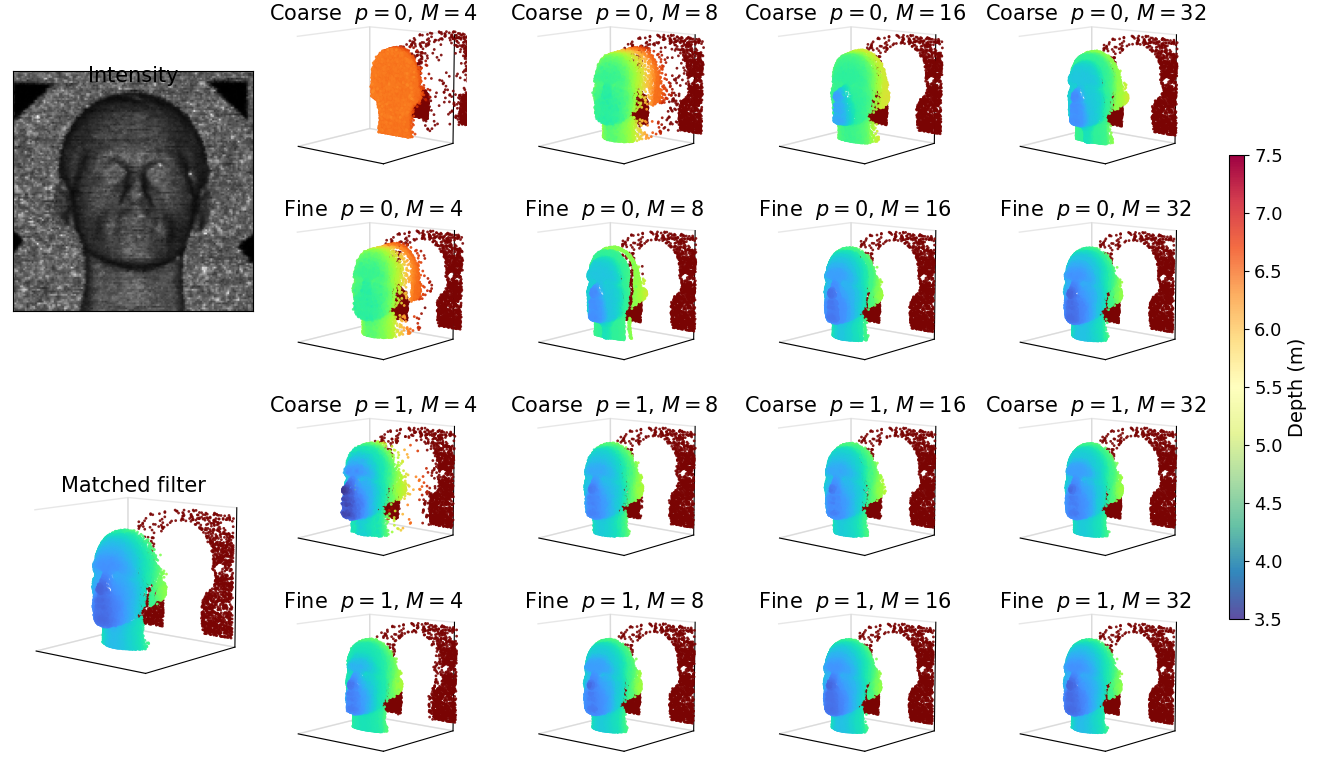}
    \caption{Reconstruction of the polystyrene-head dataset
            \cite{altmann2016lidar}.
            Acquisition: $T = 4613$ bins, bin width $\Delta_t = 19$\,ps,
            IRF FWHM = 95 ps, $141 \times 141$ pixels.
            Left column: intensity image and MF depth from the full
            $T$-bin histogram.
            Rows 1--2: CS and FS depth from the $M$-coefficient
            $p{=}0$ spline sketch.
            Rows 3--4: CS and FS depth from the $M$-coefficient
            $p{=}1$ spline sketch.
            FS reuses the same $M$-coefficient on the winning coarse basis.
            Compression ratio $T/M$ ranges from $\approx 1153\times$ at $M{=}4$ to
            $\approx 144\times$ at $M{=}32$. The face geometry is
            recovered at $M{=}8$ and visually matches the MF baseline by $M{=}16$
            for both sketch orders.}
    \label{fig:head_reconstruction}
\end{figure*}

\section{Synthetic Datasets Evaluation}
We simulate an echo pulse (FWHM = 2 ns) over a period
$T = 4096$ bins of width $80$\,ps (${\approx}\,49$ m measurement range), with $n = 500$ photons per trial. We sweep $M \in \{4, 8, 12\}$ and run 50 trials per depth on a grid of $200$ GT positions spanning 0.05 $T$ to $0.95$ $T$. Fig. \ref{fig:depth_sweep} compares estimated and ground-truth bins as the true depth is swept across the measurement range. The diagonal denotes perfect estimation, so curves lying closer to it are more accurate. At $M=4$, the FS estimates (solid) track the diagonal more closely than the CS estimates (dotted) and avoid the fluctuations visible in the CS traces in the inset. No substantial difference between CS and FS is depicted for $M>4$. We compute the CRB numerically per depth in both the CS and FS schemes. Fig. \ref{fig:rmse_crb}(a) quantifies that FS reduces the median RMSE by up to ${\sim}10\times$ over CS at $M \geq 8$, and the FS algorithm approaches the FS CRB in Fig. \ref{fig:rmse_crb}(b) for $M \geq 8$. CS and FS oscillate with period $h_{\mathrm{step}} = T/M$, peaking at basis-peak boundaries where the CF solutions of \cite{spline_sketch} switch, resolving the switching conditions. The FS CRB is roughly an order of magnitude below the CS CRB for all $M$, confirming that the gain comes from the finer window spacing $\tilde\Delta < \Delta$, which increases Fisher information about the delay and correspondingly lowers the CRB. Fig. \ref{fig:heatmaps_onepeak} maps the depth RMSE of the FS scheme across the full $(\text{FWHM},\,\text{intensity},\,\text{SBR})$ operating space. Panel (a) shows the trade-off between photon budget and background, i.e., at fixed IRF width, RMSE falls along the diagonal of increasing intensity and SBR, reaching sub-centimeter accuracy in the upper-right region. ($\geq 10^3$~photons/pixel, $\text{SBR}\geq 1$) and degrading to meter-scale errors in the low-photon and low-SBR cases. Panel (b) shows that once the photon budget is sufficient ($\gtrsim 10^3$/pixel), the algorithm is insensitive to varying FWHM, making intensity the more impactful factor. Panel (c) isolates the SBR dependence, i.e., at the 500-photon reference budget, RMSE is a few centimeters provided $\text{SBR}\geq 1$, with degradation as SBR drops below $\sim0.3$, where background photons exceed signal photons. Panels (d)-(f) repeat the same three evaluations but for peak-2 with peak-1 fixed. The results resemble peak-1 but with a $2$--$3\times$ RMSE offset, reflected in the lower amplitude, and a failure boundary shifted by roughly $3\times$ in both intensity and SBR. The shift is most visible in (d), where the meter-scale region extends into the central diagonal; in (f), noticeable failure now reaches $\text{SBR}\sim 0.3$ rather than $\sim 0.1$. The FWHM response of (b) is preserved in (e), albeit at a few-centimeter RMSE rather than the sub-centimeter of (b), reflecting the weaker second return. Overall, within the tested parameter range, CFSS operates predominantly in a photon-limited regime, with accuracy more strongly affected by photon count and SBR than by IRF width. Although the FS outperforms the CS in the single-peak case, as demonstrated above, performance degrades for the second return, which requires roughly $3\times$ the photon budget and SBR of the first to reach a comparable operating point.

\begin{figure}[!t]
    \centering
    \includegraphics[width=\columnwidth]{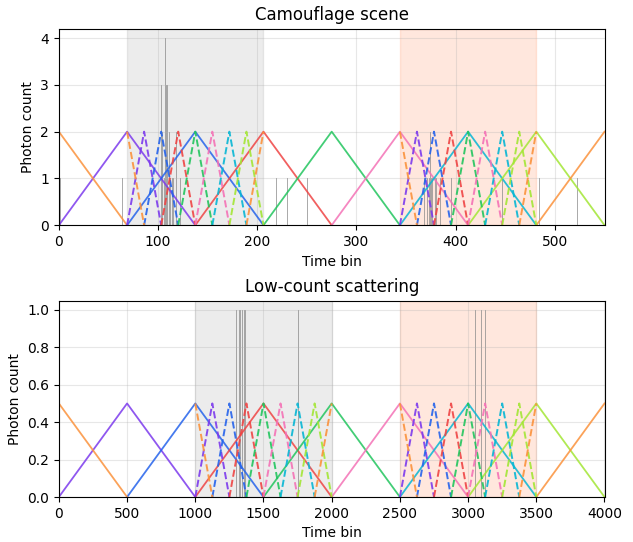}
    \caption{Representative photon histograms at pixel $(87, 38)$ from the camouflage scene \cite{tobin2018long} and low-count scattering scene \cite{shin2016computational}, with the example $M=8$ CS bases (solid triangles) and FS bases (dashed triangles). The shaded regions are the first and second FS windows identified by (\ref{eq:two-peak-two-windows}). The two histograms have photon counts of 43 and 20. The histogram lengths are 550 and 4001.}
    \label{fig:two_peak_representative_pixel}
\end{figure}

\begin{figure*}[t]
    \centering
    \includegraphics[width=\textwidth]{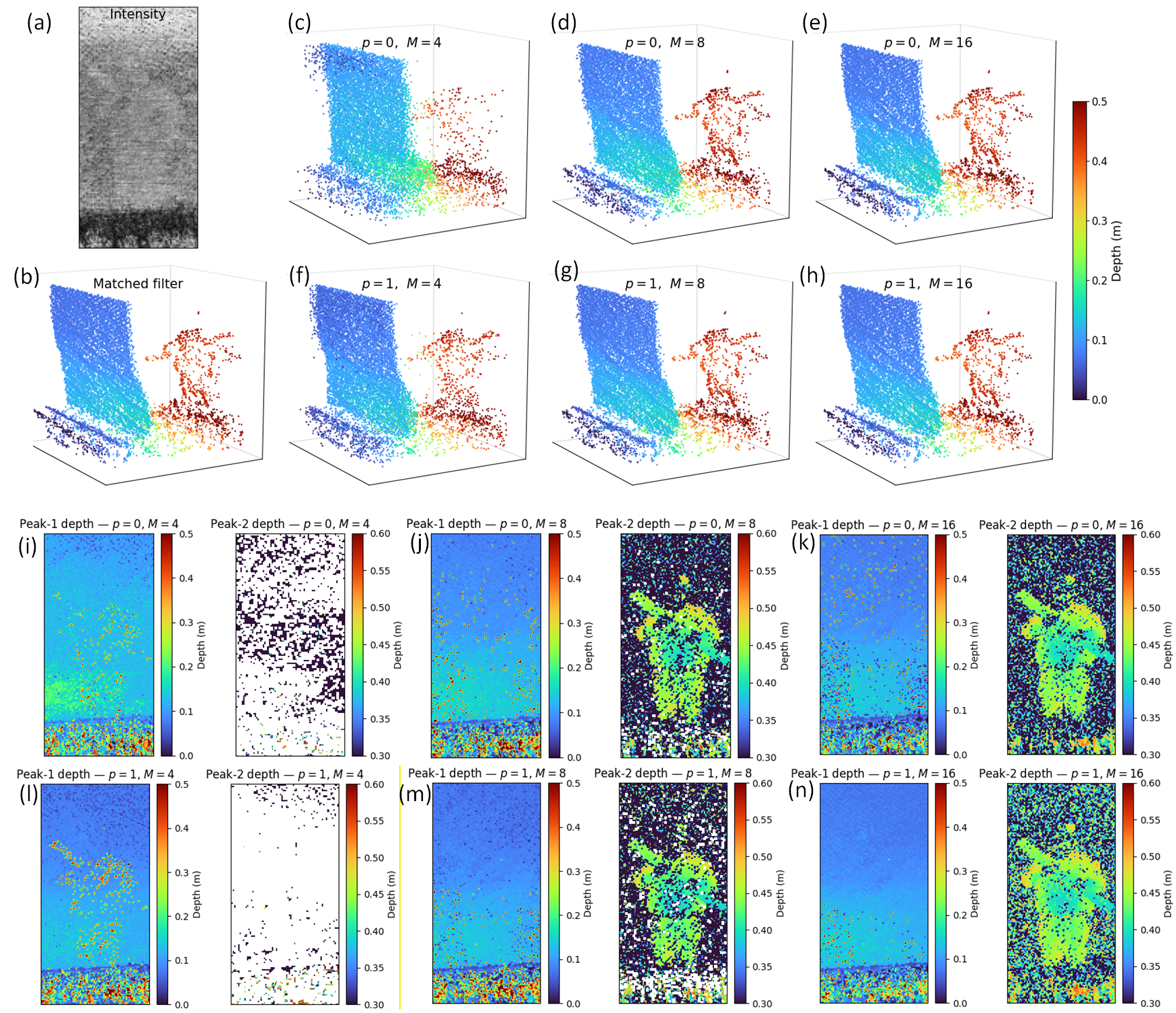}
    \caption{Real-data reconstruction of a partially occluded camouflaged
        two-return scene.
        (a) Intensity image. (b) MF reconstruction from full histograms.
        (c)--(e) Point cloud of two-peak sketch reconstructions at zero-order ($p{=}0$) for
        $M=4,8,16$.
        (f)--(h) Two-peak sketch reconstructions at first-order ($p{=}1$) for
        $M=4,8,16$.
        (i)--(n) Point cloud separate peak-1 and peak-2 depth maps for the same $M$ values,
        with $p{=}0$ in the top row and $p{=}1$ in the bottom row.
        As documented in \cite{tobin2018long}, TCSPC's bin width is 2~ps,
        and the IRF's width is 226~ps. The total distance to the target is
        $\sim$230 m, where the 1900-bin gate (0.57 m) is used as a target-focused
        depth range. White pixels mean no second peak detected.}
    \label{fig:real_two_peak_reconstruction}
\end{figure*}

Fig. \ref{fig:peak2_rmse_vs_depth} reports the depth RMSE of peak-2 for disjoint CS bins, sweeping from 15 to 44~m of a fixed peak-1 at 4.9~m, under $\alpha_1{=}0.5$, $\alpha_2{=}0.3$, FWHM = 2 ns, SBR = $8$ and only $n$ = $500$ photons per pixel. For $M{\geq}8$, the FS attains approximately 1–2 cm across the tested depth range. The $M{=}4$ curve incurs high error ($\sim10^3$~cm) with mask radius $r{=}1$; masking $\hat m_1$ and its two neighbors leaves a single unmasked bin, which acts as both the background estimate $\hat B$ and the unnormalized sketch coefficient $C_{\hat m_2}$. The threshold $\hat B + \gamma\sqrt{c_p\hat B}$ therefore always exceeds $C_{\hat m_2} = \hat B$, so at $M=4$ the disjoint-CS branch is never invoked; every pixel falls through to the shared-bin path, where peak-2 is estimated from a noise-dominated residual, producing the high RMSE. Therefore, reliable peak-2 detection requires $ M \geq 8$. We recommend $M = 8$ as it provides a good trade-off between detection accuracy and hardware overhead. Furthermore, we conduct stress tests under two-peak scenarios within the same CS bin to assess the CFSS's maximum separability. For $ M = 8$, which offers a favorable trade-off, peak-1 is centered in a CS bin, while peak-2 is swept from the left toward peak-1, as shown in Fig.~\ref{fig:two_peak_stress_test_M8}. As peak-2 moves closer to peak-1, distinguishing between the two peaks becomes progressively more difficult. Separability remains above 95\% down to a separation of $\sim7$ histogram bins, below which the two peaks can no longer be distinguished.

\begin{figure*}[t]
    \centering
    \includegraphics[width=\textwidth]{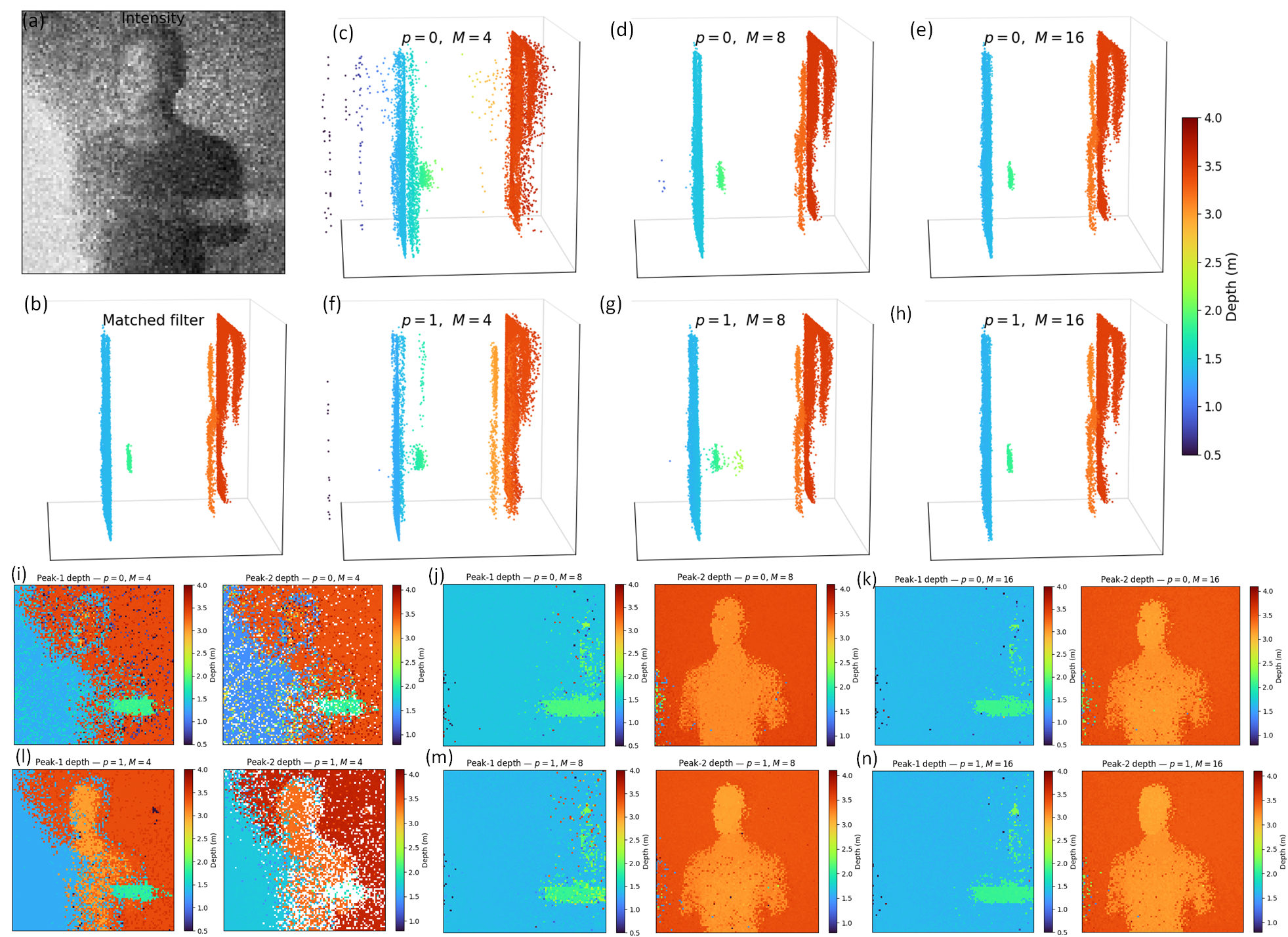}
    \caption{Low-count scattering scene \cite{shin2016computational} results.
        (a) Intensity image. (b) MF reference using the full histograms.
        (c)--(e) Two-peak sketch reconstructions at zero-order ($p{=}0$) for
        $M=4,8,16$.
        (f)--(h) Two-peak sketch reconstructions at first-order ($p{=}1$) for
        $M=4,8,16$.
        (i)--(n) Peak-1 (near wall) and peak-2 (rear-surface, mannequin and
        back wall) depth maps for the same $M$ values, with $p{=}0$ in the
        top row and $p{=}1$ in the bottom row.}
    \label{fig:MIT_scattering_two_peak}
\end{figure*}

\section{Real Datasets Evaluation}We use the experimental datasets from \cite{altmann2016lidar}, documenting the experimental setup. We include the CFSS of the zero-order spline sketch for comparison, as it performs simple coarse and fine binning in the CS and FS stages, i.e., the conventional peak-identification-and-zooming approach used in SPAD LiDAR. In subsequent figures, $p=0$ and $p=1$ denote the zero-order and first-order sketches, respectively. Fig.~\ref{fig:head_reconstruction} shows the polystyrene-head reconstruction across $M \in \{4, 8, 16, 32\}$, all of which satisfy the criterion of ~\eqref{eq:zs-regime}. All point clouds are viewed from the same angle. For $p{=}0$, naive fine-binning outperforms naive coarse-binning across different $M$, as shown in rows 1 and 2. Although our CS (row~3) at $M{=}4$ produces a clearly degraded reconstruction, the face decomposes into a few discrete depth bands and the continuous geometry of the matched filter (MF) is slightly lost; it still outperforms the $p=0$ results. FS (row 4) at $M{=}4$ recovers a more coherent face surface with only minor residual deviations from the MF reference. From $M{=}8$ to $M{=}32$, both passes are visually comparable to MF. The background wall and low-SBR outlier pixels are recovered consistently across all configurations.

We also use two experimental datasets to evaluate the two-peak scenario. These include a camouflaged, partially occluded target \cite{tobin2018long} and low-count echo responses through a flat, uniform scattering medium \cite{shin2016computational}. Fig. \ref{fig:two_peak_representative_pixel} shows a representative pixel from the scenes, along with its histogram and the CS and FS. Fig. \ref{fig:real_two_peak_reconstruction} demonstrates the reconstruction performance on the real camouflaged target. The MF result in Fig.~\ref{fig:real_two_peak_reconstruction} (b) provides a useful full-histogram reference. The proposed two-peak sketch reconstructions in Fig.~\ref{fig:real_two_peak_reconstruction} (c)--(e) and (f)---(h) reconstruct the scene geometry for $p=0, 1$, respectively. Panels (i)--(n) further illustrate the role of the two-return model, where peak-1 predominantly captures the nearer scattering layer, and peak-2 reveals responses from the target behind the camouflage. As $M$ increases from 4 to 16, the recovered first- and second-return surface becomes smoother and more consistent with the MF reference. For $M=4$, the $p=0$ scheme shows layered camouflage artifacts and a sparser point cloud for peak-2, whereas the $p=1$ scheme produces a denser and more continuous point cloud. Panels (i)--(k) and (l)--(n) show the peak-1 and peak-2 depth maps for schemes $p=0$ and $p=1$. Both schemes recover peak-1 well, with only mild distortion at $M=4$, since the two returns fall within a single coarse bin and the unresolved peak-2 aligns with the limitation shown in Fig. \ref{fig:two_peak_stress_test_M8}. The sparsest sketch ($M=4$) fails to recover peak-2, shown by the white (undetected) pixels in panel (i). In the far background, no targets lie behind the camouflage, so no second echo appears, and peak-2 detection responds only to noise. The data were acquired with a 2 ps TCSPC resolution, a 14 ns detector gate, and 226 ps FWHM system timing jitter; the target-focused processing in that work selected a 1900-bin temporal gate corresponding to an approximately 0.57 m depth.

Fig. \ref{fig:MIT_scattering_two_peak} (c)-(h) presents the reconstruction results for the low-count scattering data. Similarly, we use the full-histogram MF as the reference at $M \in \{4, 8, 16\}$. The MF resolves three depth layers: a wall near $\sim 1.5$~m, a mannequin at $\sim 3.5$~m, and the wall back at $\sim 4$~m. Apart from $M{=}4$, where coarse-binning ($p{=}0$) performs worse than $p{=}1$, results at $M{=}8, 16$ are visually comparable, with clearly separated point-cloud layers. Comparing results in panels (i) and (l), $p{=}1$ recovers a relatively clear mannequin. Increasing $M$ to $8$ and then $16$ progressively sharpens the boundaries and removes the stray points when $M=4$, nearly matching the MF reference at $M=16$. Fig. \ref{fig:MIT_scattering_two_peak} (i)---(n) separates two-peak depth maps. Peak-1 recovers the foreground scatterers, where the near wall and the mannequin's silhouette are in the lower-right, while peak-2 captures the back wall and the mannequin's body. At $M=4$, both maps are noticeably noisier, with stray pixels scattered through the background. By $M=8$, the mannequin's outline are clear in the peak-2 panel; at $M=16$, the silhouette is sharp enough to read facial geometry, and the peak-1 map cleanly isolates the foreground from the wall behind.

\section{Conclusion}
This work optimizes the sketched-LiDAR framework with finer sketch intervals that not only alleviate the high-data-rate bottleneck of modern SPAD arrays, but also resolve ToF at higher temporal resolution without increasing the per-photon arithmetic complexity. We investigate and evaluate both single- and two-peak solutions on synthetic and real datasets. The method is designed with hardware-implementation complexity in mind, facilitating efficient on-chip implementation. Future optimization can target stronger-scattering media that broaden the echo pulse, to enhance generalization. Industry demonstrators have characterized echo histogram profiles for different scattering conditions~\cite{sick_lms500_2021}, and the widened pulse may violate the performance regime of~\eqref{eq:zs-regime}. Extending the framework to such regimes is left to future work.

\section*{Acknowledgments}
We acknowledge the authors from \cite{tachella2019real}, \cite{tobin2018long}, and \cite{shin2016computational} for open-sourcing the datasets.

\IEEEtriggeratref{99}
\bibliographystyle{IEEEtran}
\bibliography{reference}

@article{shin2016photon,
  title={Photon-efficient imaging with a single-photon camera},
  author={Shin, Dongeek and Xu, Feihu and Venkatraman, Dheera and Lussana, Rudi and Villa, Federica and Zappa, Franco and Goyal, Vivek K and Wong, Franco NC and Shapiro, Jeffrey H},
  journal={Nature communications},
  volume={7},
  number={1},
  pages={12046},
  year={2016},
  publisher={Nature Publishing Group UK London}
}

@article{lindell2018single,
  title={Single-photon {3D} imaging with deep sensor fusion},
  author={Lindell, David B and O'Toole, Matthew and Wetzstein, Gordon},
  journal={ACM Trans. Graph.},
  volume={37},
  number={4},
  pages={113},
  year={2018}
}

@ARTICLE{ingle2023_equi_depth,
  author={Ingle, Atul and Maier, David},
  journal={IEEE Transactions on Pattern Analysis and Machine Intelligence}, 
  title={Count-Free Single-Photon 3D Imaging With Race Logic}, 
  year={2025},
  volume={47},
  number={9},
  pages={7292-7303},
  doi={10.1109/TPAMI.2023.3302822}}

@inproceedings{sadekar2024eui_depth,
  title={Single-photon {3d} imaging with equi-depth photon histograms},
  author={Sadekar, Kaustubh and Maier, David and Ingle, Atul},
  booktitle={European Conference on Computer Vision},
  pages={381--398},
  year={2024},
  organization={Springer}
}

@article{sun2020spadnet,
  title={SPADnet: deep {RGB}-{SPAD} sensor fusion assisted by monocular depth estimation},
  author={Sun, Zhanghao and Lindell, David B and Solgaard, Olav and Wetzstein, Gordon},
  journal={Optics express},
  volume={28},
  number={10},
  pages={14948--14962},
  year={2020},
  publisher={Optical Society of America}
}

@book{kingman1992poisson,
  title={Poisson processes},
  author={Kingman, John Frank Charles},
  volume={3},
  year={1992},
  publisher={Clarendon Press}
}

@inproceedings{gnecchi20171,
  title={A 1$\times$ 16 SiPM array for automotive {3D} imaging {LiDAR} systems},
  author={Gnecchi, Salvatore and Jackson, Carl},
  booktitle={Proceedings of the 2017 International Image Sensor Workshop (IISW), Hiroshima, Japan},
  volume={30},
  year={2017}
}

@article{shin2016computational,
  title={Computational multi-depth single-photon imaging},
  author={Shin, Dongeek and Xu, Feihu and Wong, Franco NC and Shapiro, Jeffrey H and Goyal, Vivek K},
  journal={Optics express},
  volume={24},
  number={3},
  pages={1873--1888},
  year={2016},
  publisher={Optical Society of America}
}

@article{tobin2018long,
  title={Long-range depth profiling of camouflaged targets using single-photon detection},
  author={Tobin, Rachael and Halimi, Abderrahim and McCarthy, Aongus and Ren, Ximing and McEwan, Kenneth J and McLaughlin, Stephen and Buller, Gerald S},
  journal={Optical Engineering},
  volume={57},
  number={3},
  pages={031303--031303},
  year={2018},
  publisher={Society of Photo-Optical Instrumentation Engineers}
}

@article{tachella2019real,
  title={Real-time {3D} reconstruction from single-photon {Lidar} data using plug-and-play point cloud denoisers},
  author={Tachella, Juli{\'a}n and Altmann, Yoann and Mellado, Nicolas and McCarthy, Aongus and Tobin, Rachael and Buller, Gerald S and Tourneret, Jean-Yves and McLaughlin, Stephen},
  journal={Nature communications},
  volume={10},
  number={1},
  pages={4984},
  year={2019},
  publisher={Nature Publishing Group UK London}
}

@article{hutchings2019,
  title={A reconfigurable 3-D-stacked {SPAD} imager with in-pixel histogramming for flash {LIDAR} or high-speed time-of-flight imaging},
  author={Hutchings, Sam W and Johnston, Nick and Gyongy, Istvan and Al Abbas, Tarek and Dutton, Neale AW and Tyler, Max and Chan, Susan and Leach, Jonathan and Henderson, Robert K},
  journal={IEEE Journal of Solid-State Circuits},
  volume={54},
  number={11},
  pages={2947--2956},
  year={2019},
  publisher={IEEE}
}

@article{henderson2019192,
  title={A 192 x 128 correlated {SPAD} image sensor in 40-nm {CMOS} technology},
  author={Henderson, Robert K and Johnston, Nick and Della Rocca, Francescopaolo Mattioli and Chen, Haochang and Li, David Day-Uei and Hungerford, Graham and Hirsch, Richard and Mcloskey, David and Yip, Philip and Birch, David JS},
  journal={IEEE Journal of Solid-State Circuits},
  volume={54},
  number={7},
  pages={1907--1916},
  year={2019},
  publisher={IEEE}
}

@article{zhang201830,
  title={A 30-frames/s, 252 x 144 {SPAD} flash {LiDAR} with 1728 dual-clock 48.8-ps {TDC}s, and pixel-wise integrated histogramming},
  author={Zhang, Chao and Lindner, Scott and Antolovi{\'c}, Ivan Michel and Pavia, Juan Mata and Wolf, Martin and Charbon, Edoardo},
  journal={IEEE Journal of Solid-State Circuits},
  volume={54},
  number={4},
  pages={1137--1151},
  year={2018},
  publisher={IEEE}
}

@article{binning,
  title={High-speed {3D} sensing via hybrid-mode imaging and guided upsampling},
  author={Gyongy, Istvan and Hutchings, Sam W and Halimi, Abderrahim and Tyler, Max and Chan, Susan and Zhu, Feng and McLaughlin, Stephen and Henderson, Robert K and Leach, Jonathan},
  journal={Optica},
  volume={7},
  number={10},
  pages={1253--1260},
  year={2020},
  publisher={Optical Society of America}
}

@inproceedings{sliding,
  title={A reconfigurable QVGA/Q3VGA Direct time-of-flight {3D} imaging system with on-chip depth-map computation in 45/40 nm 3D-stacked {BSI SPAD CMOS}},
  author={Stoppa, David and Abovyan, Sargis and Furrer, Daniel and Gancarz, Radoslaw and Jessenig, Thomas and Kappel, Robert and Lueger, Manfred and Mautner, Christian and Mills, Ian and Perenzoni, Daniele and others},
  booktitle={Proc. Int. Image Sensor Workshop},
  pages={53--56},
  year={2021}
}

@inproceedings{zooming,
  title={A 16.5 giga events/s 1024$\times$ 8 {SPAD} line sensor with per-pixel zoomable 50ps-6.4 ns/bin histogramming {TDC}},
  author={Erdogan, Ahmet T and Walker, Richard and Finlayson, Neil and Krstajic, Nikola and Williams, Gareth OS and Henderson, Robert K},
  booktitle={2017 Symposium on VLSI Circuits},
  pages={C292--C293},
  year={2017},
  organization={IEEE}
}

@inproceedings{lin2024spiking,
  title={Spiking neural networks for active time-resolved {SPAD} imaging},
  author={Lin, Yang and Charbon, Edoardo},
  booktitle={Proceedings of the IEEE/CVF Winter Conference on Applications of Computer Vision},
  pages={8147--8156},
  year={2024}
}

@article{macleanTDCless,
  title={{TDC}-less direct time-of-flight imaging using spiking neural networks},
  author={MacLean, Jack Iain and Stewart, Brian D and Gyongy, Istvan},
  journal={IEEE Sensors Journal},
  volume={24},
  number={20},
  pages={33838--33846},
  year={2024},
  publisher={IEEE}
}

@ARTICLE{Histless_lidar,
  author={Tontini, Alessandro and Mazzucchi, Sonia and Passerone, Roberto and Broseghini, Nicolò and Gasparini, Leonardo},
  journal={IEEE Sensors Journal}, 
  title={Histogram-Less LiDAR Through {SPAD} Response Linearization}, 
  year={2024},
  volume={24},
  number={4},
  pages={4656-4669},
  doi={10.1109/JSEN.2023.3342609}}

@ARTICLE{Histless_lidar_auto,
  author={Feng, Lichen and Liu, Yimeng and Li, Dong and Bao, Yaoqi and Ma, Rui and Zhu, Zhangming},
  journal={IEEE Sensors Journal}, 
  title={Robust Compressive Histogramming Based on Autoencoder for {SPAD} Direct {ToF} {LiDAR} Covering Challenging Scenarios}, 
  year={2025},
  volume={25},
  number={14},
  pages={27701-27711},
  doi={10.1109/JSEN.2025.3575784}}

@article{istvan_4x4,
  title={A direct time-of-flight image sensor with in-pixel surface detection and dynamic vision},
  author={Gyongy, Istvan and Erdogan, Ahmet T and Dutton, Neale AW and Mart{\'\i}n, Germ{\'a}n Mora and Gorman, Alistair and Mai, Hanning and Della Rocca, Francesco Mattioli and Henderson, Robert K},
  journal={IEEE Journal of Selected Topics in Quantum Electronics},
  volume={30},
  number={1: Single-Photon Technologies and Applications},
  pages={1--11},
  year={2023},
  publisher={IEEE}
}

@ARTICLE{sketch_lidar,
  author={Sheehan, Michael P. and Tachella, Julián and Davies, Mike E.},
  journal={IEEE Transactions on Computational Imaging}, 
  title={A Sketching Framework for Reduced Data Transfer in Photon Counting {Lidar}}, 
  year={2021},
  volume={7},
  number={},
  pages={989-1004},
  doi={10.1109/TCI.2021.3113495}}

@ARTICLE{spline_sketch,
  author={Sheehan, M. P. and Tachella, J. and Davies, M. E.},
  journal={IEEE Transactions on Computational Imaging}, 
  title={Spline Sketches: An Efficient Approach for Photon Counting {Lidar}}, 
  year={2024},
  volume={10},
  number={},
  pages={863-875},
  doi={10.1109/TCI.2024.3404652}}

@article{spline_fpga,
  title={{Fpga} implementation of sketched {Lidar} for a 192 x 128 {SPAD} image sensor},
  author={Zang, Zhenya and Davies, Mike and Gyongy, Istvan},
  journal={arXiv preprint arXiv:2602.10837},
  year={2026}
}

@article{altmann2016lidar,
  title={{Lidar} waveform-based analysis of depth images constructed using sparse single-photon data},
  author={Altmann, Yoann and Ren, Ximing and McCarthy, Aongus and Buller, Gerald S and McLaughlin, Steve},
  journal={IEEE Transactions on Image Processing},
  volume={25},
  number={5},
  pages={1935--1946},
  year={2016},
  publisher={IEEE}
}

@misc{sick_lms500_2021,
  author       = {{SICK}},
  title        = {{SICK LMS500-20000 PRO HR Indoor Laser Scanner}},
  howpublished = {\url{https://www.generationrobots.com/en/401697-sick-lms500-20000-pro-hr-indoor-laser-scanner.html}},
  note         = {}
}

\end{document}